\documentclass[12pt]{iopart}
\usepackage{epsf}
\usepackage{graphicx}
\begin{document}

\title{Electronic structure of Diluted Magnetic Semiconductors $Ga_{1-x}Mn_{x}N$ and
$Ga_{1-x}Cr_{x}N$} 

\author{Nandan Tandon$^1$, G P Das$^2$ and Anjali Kshirsagar$^1$\footnote{Author 
to whom any correspondence should be addressed.}}
\address{$^1$Department of Physics,
University of Pune,
Pune 411 007, India\\
$^2$Department of Material Science, Indian Association for the Cultivation of Science,
Jadavpur, Kolkata, 700032, India}

\ead{anjali@physics.unipune.ernet.in}

%\date{\today}

\begin{abstract}
We have undertaken a study of diluted magnetic semiconductors $Ga_{1-x}Mn_{x}N$ and $Ga_{1-x}Cr_{x}N$ 
with $x=0.0625, 0.125$, using the all electron linearized augmented plane wave method (LAPW) 
for different configurations of Mn as well as Cr. We study 
four possible configurations of the impurity in the wurtzite GaN structure to predict 
energetically most favorable structure within the 32 atom supercell and conclude that the
near-neighbor configuration has the lowest energy. We have also analyzed the ferro-magnetic
as well as anti-ferromagnetic configurations of the impurity atoms. The density of states as well as
bandstructure indicate half metallic state for all the systems. $T_c$ has also been estimated
for the above systems.
\end{abstract}


\maketitle

\begin{section}{Introduction}
Gallium nitride is one of the most promising materials among the diluted magnetic semiconductor
(DMS) material for application in spintronics. By doping transition metal (TM) atoms, Mn or Cr,
local magnetic moment are introduced in 
semiconductor which mediate ferromagnetically.  (Ga,Cr)N based DMS was predicted
to show high $T_c$~\cite{sato} for high enough concentration of Cr
and further Hashimoto {\it et al.}~\cite{cr1} 
observed that (Ga,Cr)N based DMS grown by ECR molecular beam epitaxy showed $T_c$ above $400$K. 
Cr$^+$-implanted GaN, studied by photoluminescence and superconducting quantum interference device
(SQUID) reveal that the implanted Cr$^+$ incorporates substitutionally at Ga site and the
ferromagnetic order is retained upto $300$K~\cite{cr2}. Takeuch {\it et al.}~\cite{take}. 
have reported a systematic
study of changes in the occupied and unoccupied N-partial density of states (DOS) and 
confirm the wurtzite N $2-p$ DOS and substitutional doping of Cr into Ga sites 
using SXES and XAS. Recently, ferromagnetism above $900$K was reported in Cr-GaN thin films~\cite{liu}. 
Theoretically it was predicted that the ferromagnetic (FM) interaction
in (Ga,Mn)N may be retained upto room temperature~\cite{diet}. The initial reports of high $T_c$ in 
(Ga,Mn)N were followed by controversial results where the reported $T_c$ varied between $20$K -
$940$K~\cite{reed, thal, sono, ando}. Zajac and coworkers observed Mn ions in Ga$_{1-x}$Mn$_{x}$N ($x < 0.1$)
crystals coupled anti-ferromagnetically (AFM)~\cite{zaja}. Electronic structure
and magnetic properties of zinc blende Ga$_{1-x}$Mn$_{x}$N for several values of $x$ 
with varied spatial
distribution of dopant atoms to understand the magnetic interaction for explanation of FM-AFM
competition is discussed by Uspenskii {\it et al.}~\cite{uspen} where the calculations were
done using the tight binding LMTO method in the local spin density approximation. 
Sanyal and Mirbt~\cite{mirb} have
studied Mn doped GaAs and GaN DMS using the {\it ab-initio} plane wave code (VASP) within density
functional theory (DFT). They have determined the interatomic exchange interactions by
substituting Mn in various positions in the unit cell and have attributed the origin of 
ferromagnetism in (Ga,Mn)N to double-exchange mechanism involving the hopping of Mn$-d$ electrons.
Raebiger {\it et al.}~\cite{reib} used the full potential linearized augmented plane wave (FP-LAPW) method to
investigate the interplay between clustering and exchange coupling in magnetic semiconductor
Ga$_{1-x}$Mn$_{x}$As. They have studied all possible arrangements of the two Mn atoms on Ga 
sublattice for $x \sim 6 \%$ and found that clustering of Mn atoms at near neighbour Ga sites is
energetically preferred. Our analysis of the wurtzite GaN doped with Mn or Cr is motivated by
the latter study.
\end{section}

\begin{section}{Method and Computational Details}
We have employed the spin-polarized Linearized Augmented Plane Wave Method (FP-LAPW)
as implemented in the WIEN2K package~\cite{wien} with the Generalized Gradient 
Approximation (GGA) for the exchange-correlation potential proposed by Perdew,
Burke and Ernzerof (PBE96)~\cite{pbe96}.
This is state-of-the-art electronic structure
method, which does not use any shape approximation for the potential, to solve the 
Kohn-Sham type of equations self-consistently.

GaN normally occurs in the wurtzite structure with lattice constants $a=3.19$\AA~and
$c=5.19$\AA, giving $c/a$ ratio of 1.62. Each Ga is tetrahedrally bonded 
to N atoms at an average distance of
$1.95$\AA~and each N in turn is surrounded by four Ga neighbors. The calculations for
DMS were performed within a 32 atom supercell, constructed from $2\times2\times2$ standard
unit cell of wurtzite structure wherein the dopant is substituted at various cation sites,
since it has been shown that the formation energy for interstitial Mn doping is higher than
substitutional doping~\cite{mirb}.
The supercell approach is used to restrict the dopant concentration to a small value, 
which is of interest for studying magnetic properties of the system, without altering
the original underlying lattice structure. Our interest was in observing the changes
in the electronic structure of the DMS with respect to the possible different 
geometries of the dopants within the
host semiconductor. Self-consistent electronic structure 
calculations were performed using the
APW + local orbitals (lo) basis set for the valence and semi-core electrons with
$R_{MT}*K_{max} = 7$, $l_{max}=10$ and total energy convergence of $0.00001 Ry.$
The Muffin-Tin radii for Ga, Mn and Cr were kept at $2.0$\AA~ and that for N at $1.6$\AA.
Spin-polarized calculations were carried out to observe the effect of spin-splitting and
to calculate the on-site magnetic moment at TM site.

We have studied wurtzite GaN doped with one TM atom impurity, which is $\sim 6.25 \%$ doping
and two identical TM atoms in the $32$ atom unit cell amounting to $\sim 12.5 \%$ doping. 
To simulate different surroundings for the transition metal (TM) atoms 
we have spanned certain geometries within the $32$ atom unit cell wherein
the distance between the dopants is varied. 
In the case of the single impurity substitution, the nearest distance between two TM atoms
is $6.38$\AA~ in plane and $10.38$\AA~along the $c$-axes. We have studied four 
different geometries of two TM atom substitutions at $3.19$\AA, near neighbor (nn), $4.5$\AA~, 
$5.19$\AA~ and $6.08$\AA~ separations. When two near neighbor (nn)
Ga atoms are substituted by TM atoms the in-plane TM-TM atoms distance is $3.19$\AA~ and
along the $c$-axes it is $10.38$\AA. For the second case the out-of-plane distance is $4.5$\AA~
and the in-plane distance between the dopants is $6.3$\AA~.
In the third case, two Ga atoms lying one above the other, along the
$z$-axes, separated by a distance of $5.19$\AA~ are substituted by TM atoms and the in-plane
TM atoms are at $6.38$\AA~. The last case is such that the in-plane separation
($6.38$\AA) and out of plane separation ($6.09$\AA) between the TM atoms is comparable.
For estimating the magnetically favorable system, the spins of
the dopants are aligned along the same direction, corresponding to the FM 
configuration, and aligned in the opposite directions corresponding to AFM
configuration. The self consistency was achieved on a mesh of $5\times5\times2$ k-points. 

Structural relaxation for the TM site and the nn N sites was carried out to observe 
changes in the bond lengths between TM and the first shell of N atoms. 
Very small change ($\sim 3-4 \%$)
was observed in the bond lengths and no significant changes were seen in the
band structure in agreement with the earlier reported results~\cite{kron}. 
Thus the calculations reported here are for systems without allowing any relaxation.
\end{section}

\begin{section}{Results}
In the wurtzite GaN semiconductor, each Ga (N) is tetrahedrally bonded to 4 N (Ga) atoms.
Pure GaN is a direct band gap semiconductor with top of the valence band consisting of N
{\it p}-states and the bottom of the conduction band having Ga {\it sp}-character. The 
Ga {\it d} levels are deep and do not take part in the bonding. Thus they
are treated as core states. The band gap of GaN, which is underestimated by 
density functional theory within the approximation used for the exchange correlation
energy functional, is $\sim2$eV. The experimentally determined band gap of 
undoped GaN is $\sim3.4$eV. Das {\it et al.}~\cite{gp1} have shown that, for 
Mn atoms to couple ferromagnetically, they need to be kept apart by more than 
the critical distance of $2.5$\AA. Similar calculations on clusters of (GaN)Cr 
indicate that the critical Cr-Cr distance is $2.7$\AA~\cite{gp2}. 
In all our calculations the distance between the dopants was greater than the corresponding
critical distances.

\begin{subsection}{Mn doped systems}
Localized magnetic moments are introduced within the GaN system by substituting the cations with
TM impurity atom(s). Mn atom with $4s^2$ and $3d^5$ electrons in the valence region replaces
Ga atom with valency $4s^2 4p^1$. On substitution Mn atoms contributing five $d$ levels per atom
are thus expected to contribute to the observed magnetic moment. 
\begin{figure}[h]
\hskip 1.0in
\epsfxsize 2.0in
\epsffile {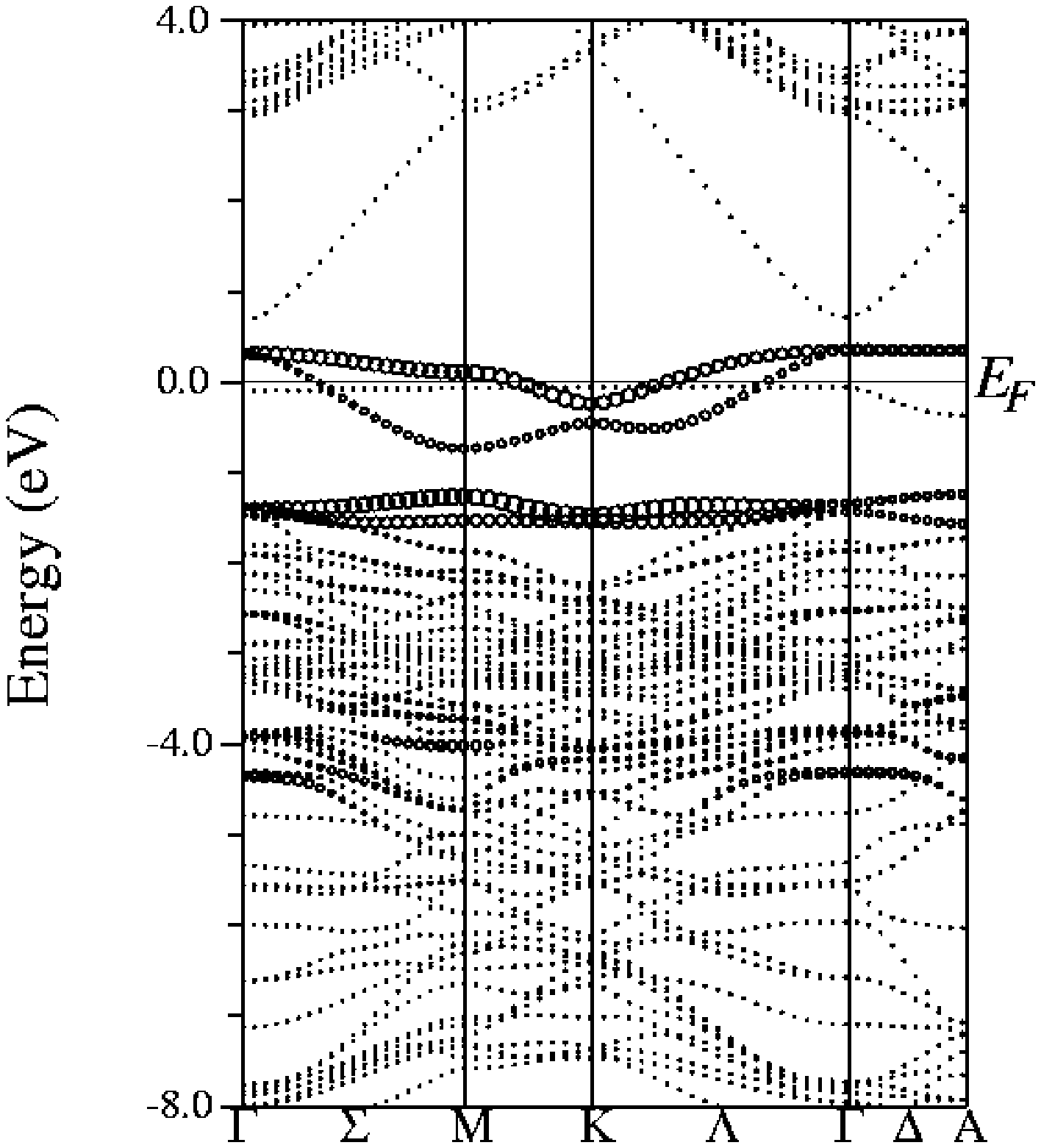}
\hskip 0.2in
\epsfxsize 2.0in
\epsffile {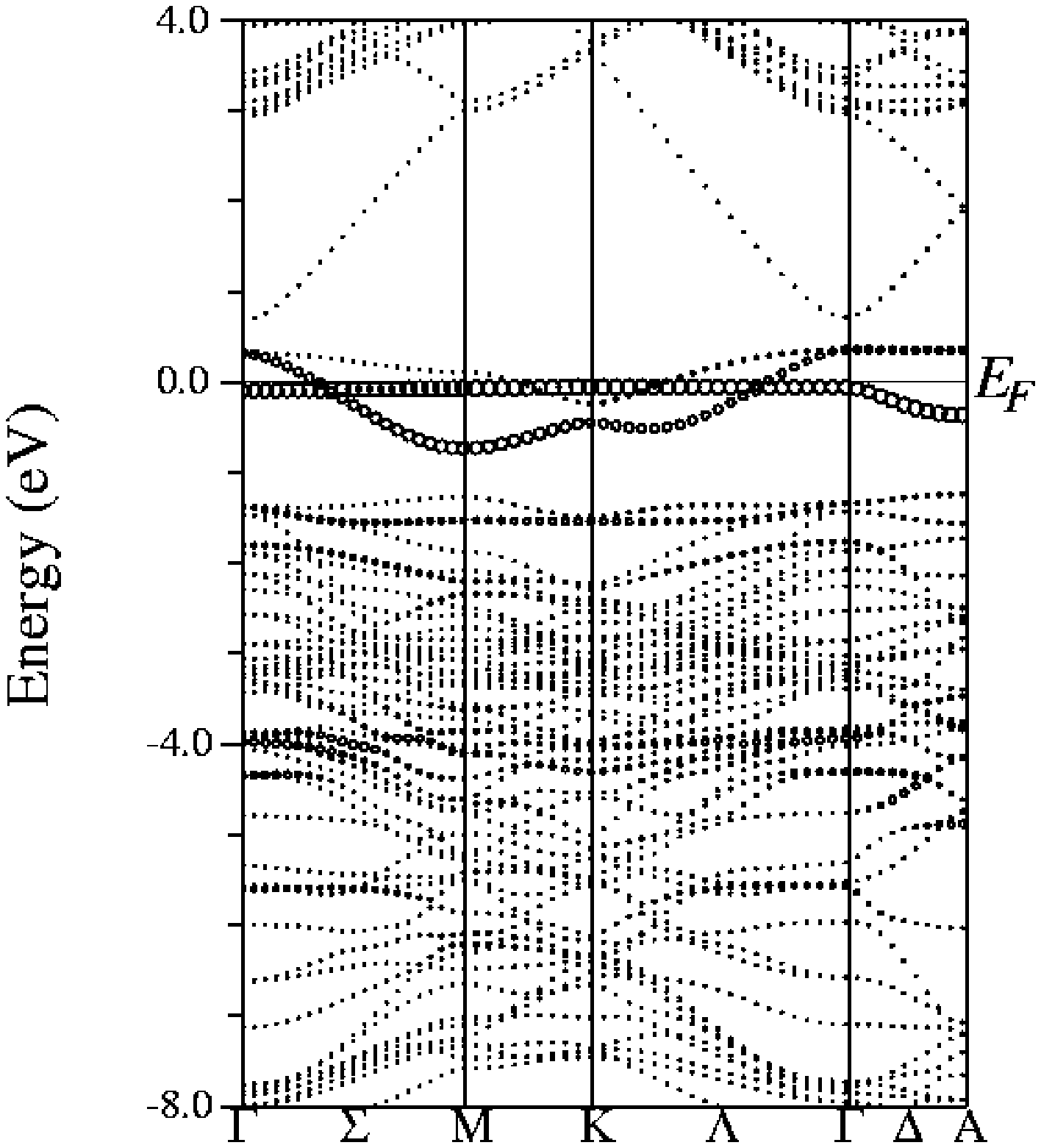}\\
\vskip 0.02cm
\hskip +2.1in {(a)} \hspace{1.8in} {(b)}
\caption{\label{mnet2}Mn-projected majority spin (a) $d_{t_2}$ and (b) $d_{e}$ bands in $Ga_{15}MnN_{16}$.}
%\caption{\label{mnet2}Mn-projected $d_{t_2}$ and $d_{e}$ majority bands in $Ga_{15}MnN_{16}$.}
\hskip 0.2in
\epsfxsize 1.5in
\end{figure}
Since three of the valence electrons from Mn go into compensating the three
electron states of substituted Ga, one hole per Mn is introduced into the system. 
Figure~\ref{mnet2} shows the Mn-projected $d_{t_2}$ and $d_e$ majority spin
electronic structure for $Ga_{15}MnN_{16}$. 
The Mn-$d$
states lie at the top of the valence band and cross the $\epsilon_F$ in some places.
These are split into $d_e$ and $d_{t_2}$ states, the $d_{t_2}$ level is two thirds filled and the
$d_e$ is almost occupied. The minority spin levels are empty and lie above the $\epsilon_F$ 
indicating $100 \%$ spin-polarized states. The Mn induced states lie in the gap 
region of GaN.
The top of the valence band in GaN is composed of the N-$p$ levels and the unique properties, 
particularly the half metallic state of DMS, thus arise from the TM {\it d} and host {\it p} 
interactions that couple the two subsystems.  

\begin{figure}[h]
\centering
\includegraphics[height=8.5cm,angle=270] {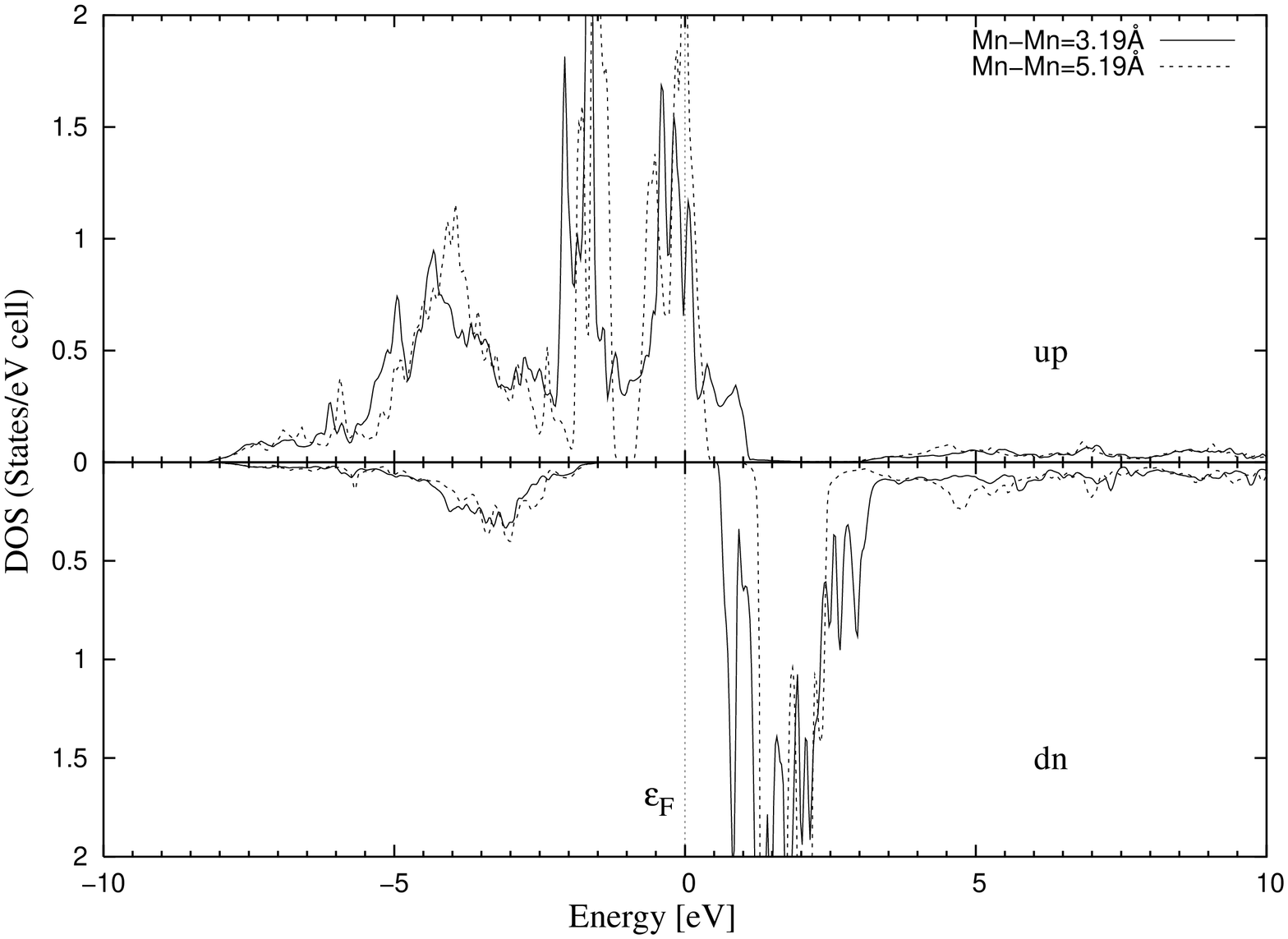}\\
%\hskip +0.4in \hspace{0.3in} {(a)}\\
{(a)}\\
\includegraphics[height=8.5cm,angle=270] {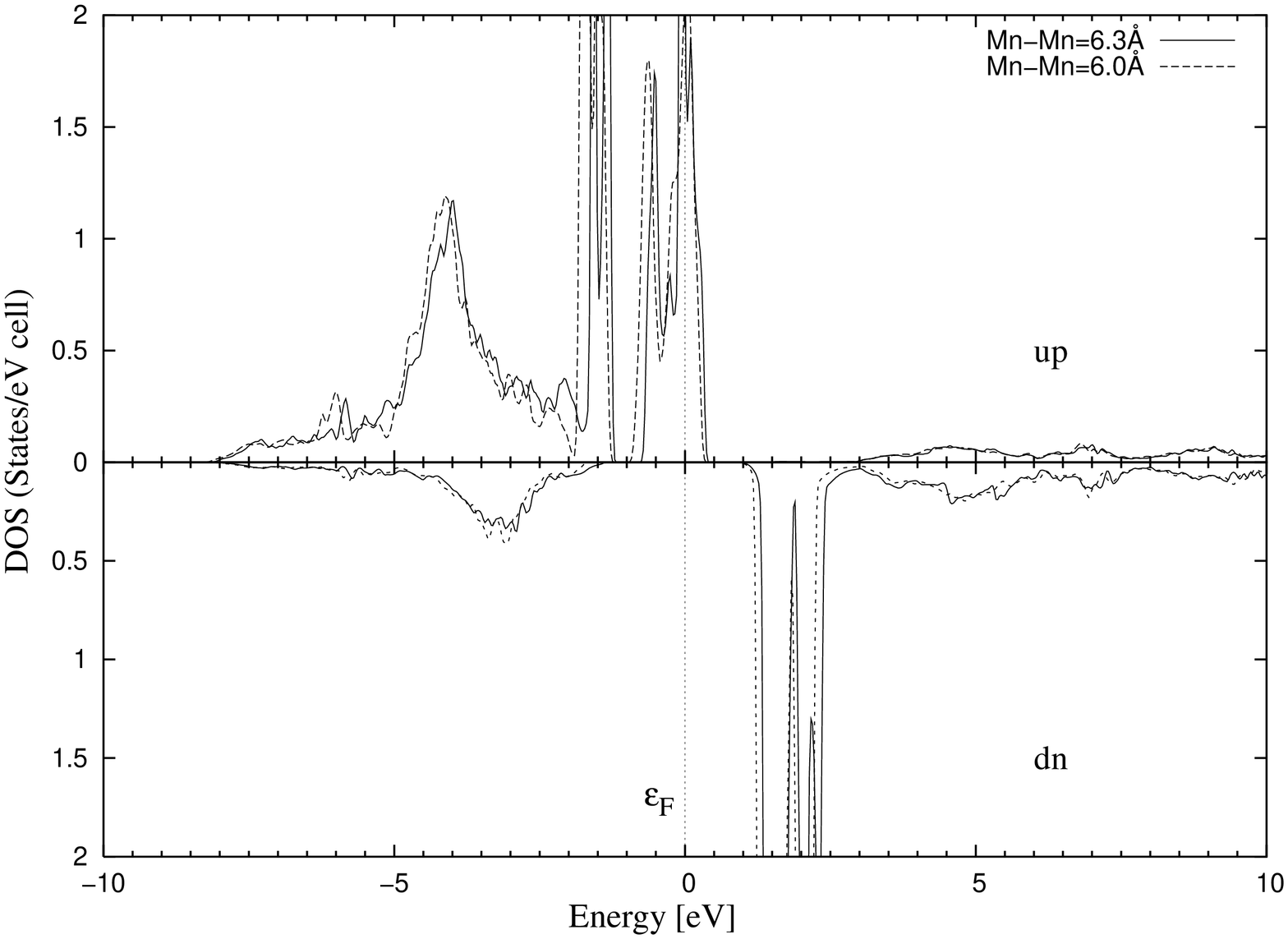}\\
%\hskip +0.4in \hspace{0.3in} {(b)}
{(b)}
\caption{\label{mndos}Mn-projected $d$-DOS in $Ga_{14}Mn_2N_{16}$ (a) with Mn-Mn distance equal 
to $3.19$\AA~ and $5.19$\AA~ and (b) $6.0$\AA~ and $6.3$\AA. The upper and lower panels represent
the majority and minority spin DOS respectively.}
\end{figure}
In order to understand the variation of exchange interaction among the TM impurity with the
distance between the TM atoms, the concentration of Mn atoms was increased
to $12.5 \%$, equivalent to introducing $2$ Mn atoms in the supercell.
Self consistent calculations were carried out for two
different magnetic configurations of Mn electrons in which the electrons are
parallel or antiparallel corresponding to FM or AFM configuration. For all the 
geometries of the dopants, systems, as described in section II, with FM configuration of the Mn
atoms were found to have lower energy. Since the Mn-$d$ levels are responsible for observed half metallic
behavior, a comparison of Mn-$d$ DOS in various geometries is shown in 
figure~\ref{mndos}, (a) for separations $3.19$\AA~ and $5.2$\AA~ and 
(b) for $6.0$\AA~ and $6.3$\AA. 
Here the half metallic state is evident in all the cases. The TM-TM distance of $6.3$\AA~
corresponds to single TM doping ($6.25\%$).

The Mn $d-$DOS is broad for substitution at nn distance. In all the other cases the 
band is split indicating that the majority spin $d$-bands of the two Mn atoms at nn overlap. On
increasing Mn-Mn distance, $d$-band splitting takes place implying a reduction in the interaction
between the TM atoms. It may be noted that the minority spin conduction
band overlaps with the  majority spin band for nn substitution. A gap of
$\sim 0.5eV$ is present between the minority spin conduction band and majority spin band for
Mn-Mn distance greater than nn.
The minority spin valence band as well as conduction band is far apart from the $\epsilon_F$ thus retaining
the highly spin polarized state also seen in the single Mn doped $Ga_{14}MnN_{16}$ system.
The down spin gap is $2eV$ for nn configuration and increases to $2.5eV$ at
larger separations. On increasing the distance between the Mn atoms, splitting of d-level
increases. This is consistent with the observation that in single impurity doping, TM-TM 
atom distance is $6.0$\AA~ and splitting of the Mn-$d$ band is larger as seen in figure~\ref{mndos}(b).
The magnetic moment at Mn-site does not depend on the distance
between the dopant atoms and has a value $3.34 \mu_B$ for all the geometries as indicated in
table~\ref{table1}. %in agreement with Sanyal {\it et al.}~\cite{mirb1}.

\begin{table}[h]
\caption{\label{table1}Magnetic moment($\mu_B$) in various Mn-doped systems.
Total$_{mom}$ indicates mag. mom./unit cell, dopant$_{mom}$ is the
mag. mom. at dopant site and N$_{mom}$ is the average mag. mom. at
nn N sites.}
%\vskip 0.5in
%\begin{ruledtabular}
\begin{indented}
\item[]\begin{tabular}{@{}llll}
%\begin{tabular}{c c c c }
\br
System & Total$_{mom}$ & Dopant$_{mom}$ & N$_{mom}$\\
%       & mom.($\mu_B$) & at dopant site($\mu_B$) & at nn-N site($\mu_B$)\\
\hline
$Ga_{15}MnN_{16}$     & 4.00 & 3.33 & 0.001\\
\hline
$Ga_{14}Mn_{2}N_{16}$ & 8.00 & 3.34 & 0.001\\
Mn-Mn  = $3.19$ \AA &   &  & \\
\hline
$Ga_{14}Mn_{2}N_{16}$ & 8.00 & 3.34 & 0.005\\
Mn-Mn  = $4.5$ \AA &   &  & \\
\hline
$Ga_{14}Mn_{2}N_{16}$ & 8.00 & 3.33 & 0.006\\
Mn-Mn  = $5.19$ \AA &   &  & \\
\hline
$Ga_{14}Mn_{2}N_{16}$ & 8.00 & 3.34 & 0.005 \\
Mn-Mn  = $6.0$ \AA &   &  & \\
\br
%\hline
\end{tabular}
\end{indented}
%\end{ruledtabular}
\end{table}

The presence of localized moment influences the near neighbor N atoms within the GaN
system, such that DOS of the nn N-atoms around the impurity atom
becomes as shown in figure~\ref{ndos}(a). Due to the $p-d$ interaction, induced states 
are seen on N atoms. The magnitude of the
induced states is maximum at nn N atoms and decreases as the distance from the TM atom 
increases. This again indicates
a localized nature of the TM states. Figure~\ref{ndos}(b) shows the 
spin charge density (SCD) in a plane containing three nn N atoms for single Mn doping.
The plot shows that the SCD on the N atoms lying above
the Mn atom is negative whereas it is positive for the N atom lying below the Mn atom.
Average magnetic moment on the nn N atoms is positive as shown in table~\ref{table1} for
all the different geometries.

\begin{figure}[h]
\centering
\includegraphics[height=6.5cm] {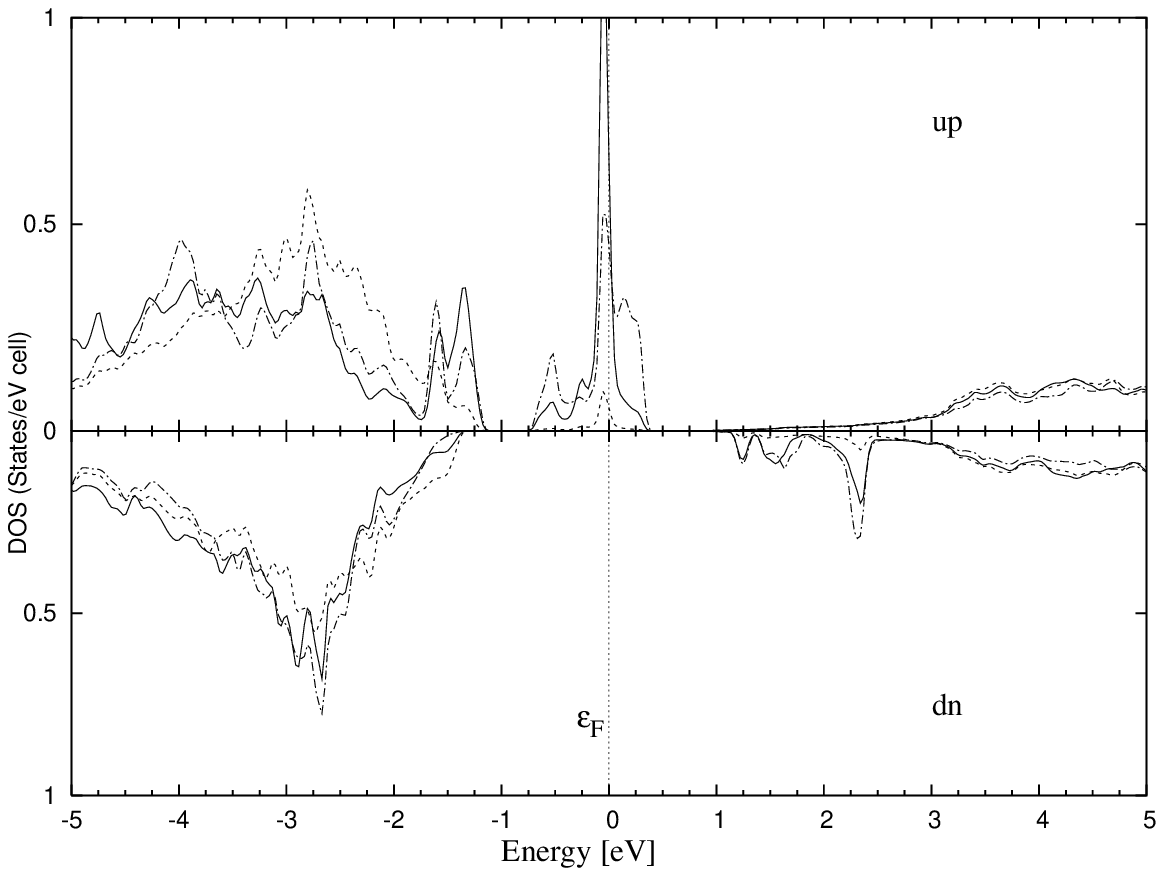}\\
{(a)}\\
\vskip 0.2cm
\includegraphics[height=5.0cm,width=5cm] {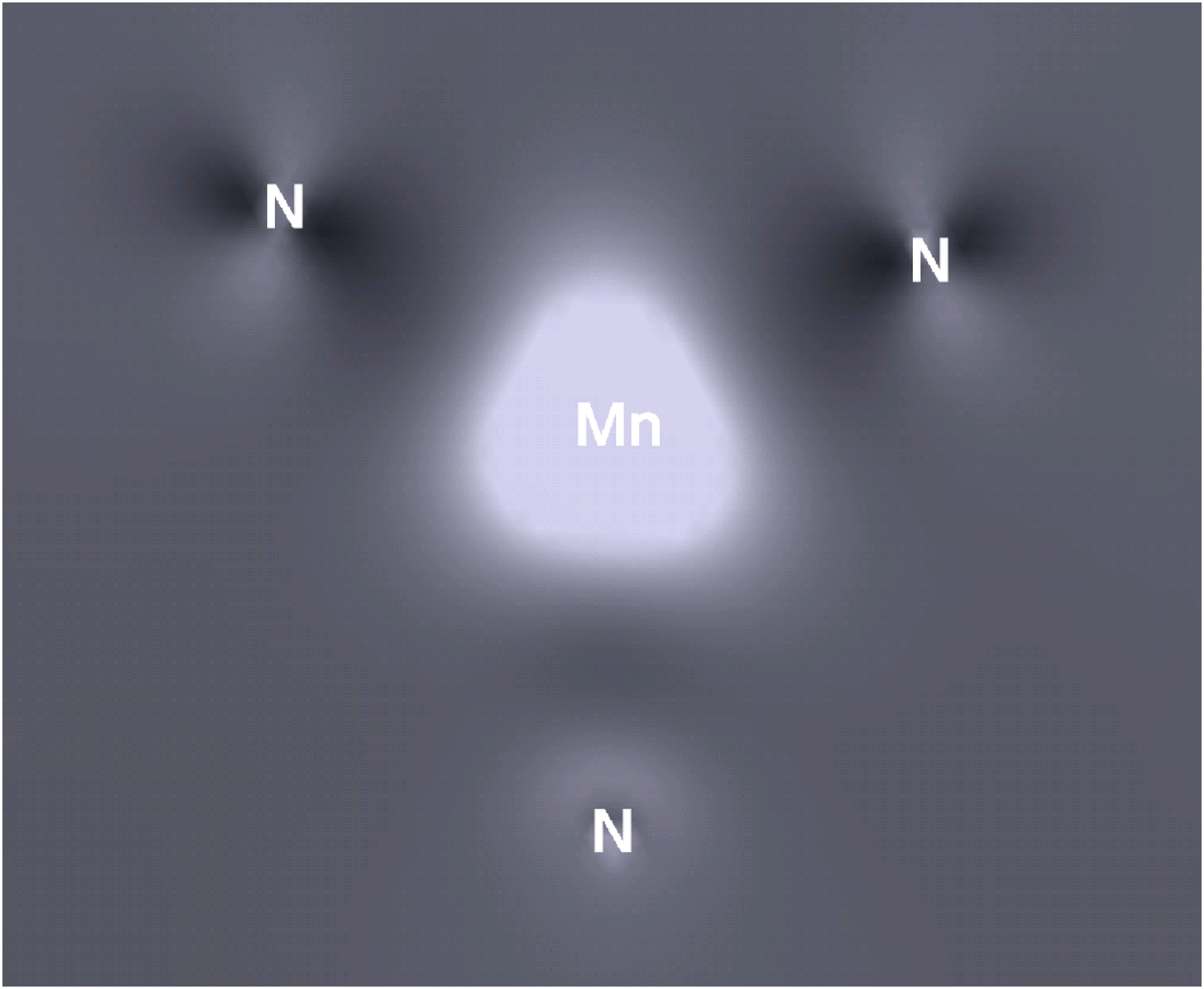}
\includegraphics[height=5.0cm,width=2cm] {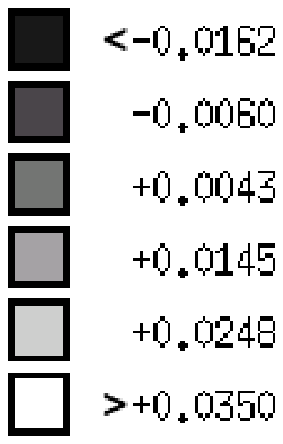}\\
{(b)}
\caption{\label{ndos}(a) Variation in partial-DOS at 3 N-sites in $Ga_{15}MnN_{16}$. 
The solid lines: nn-N along $c-$axis. Dotted lines:nn-N in plane. Dashed-dotted lines:
next nn-N (b) SCD plot in a plane of three nn N atoms.}
\end{figure}
\end{subsection}

\begin{subsection}{Cr doped systems}
Electronic structure calculation for substitutional doping of Cr 
in the GaN system was also done and is analyzed in a similar fashion.
For each Cr doped in the $32$ atom supercell, equivalent to 
$6.25 \%$ doping, there are five spin up $d$-states which are introduced in the GaN
band gap. Since the Cr atom has $4s^2 3d^4$ valence electrons, only three of 
the electron states out of the five $3-d$ levels are occupied, 
thus creating two hole states per Cr substitution.
Single Cr doping into the $32$ atom supercell at cation site results
in the Cr-$d$ levels appearing in the band gap of the semiconductor host as seen in 
figure~\ref{cret2}.
\begin{figure}[h]
\hskip 1.0in
\epsfxsize 2.0in
\epsffile {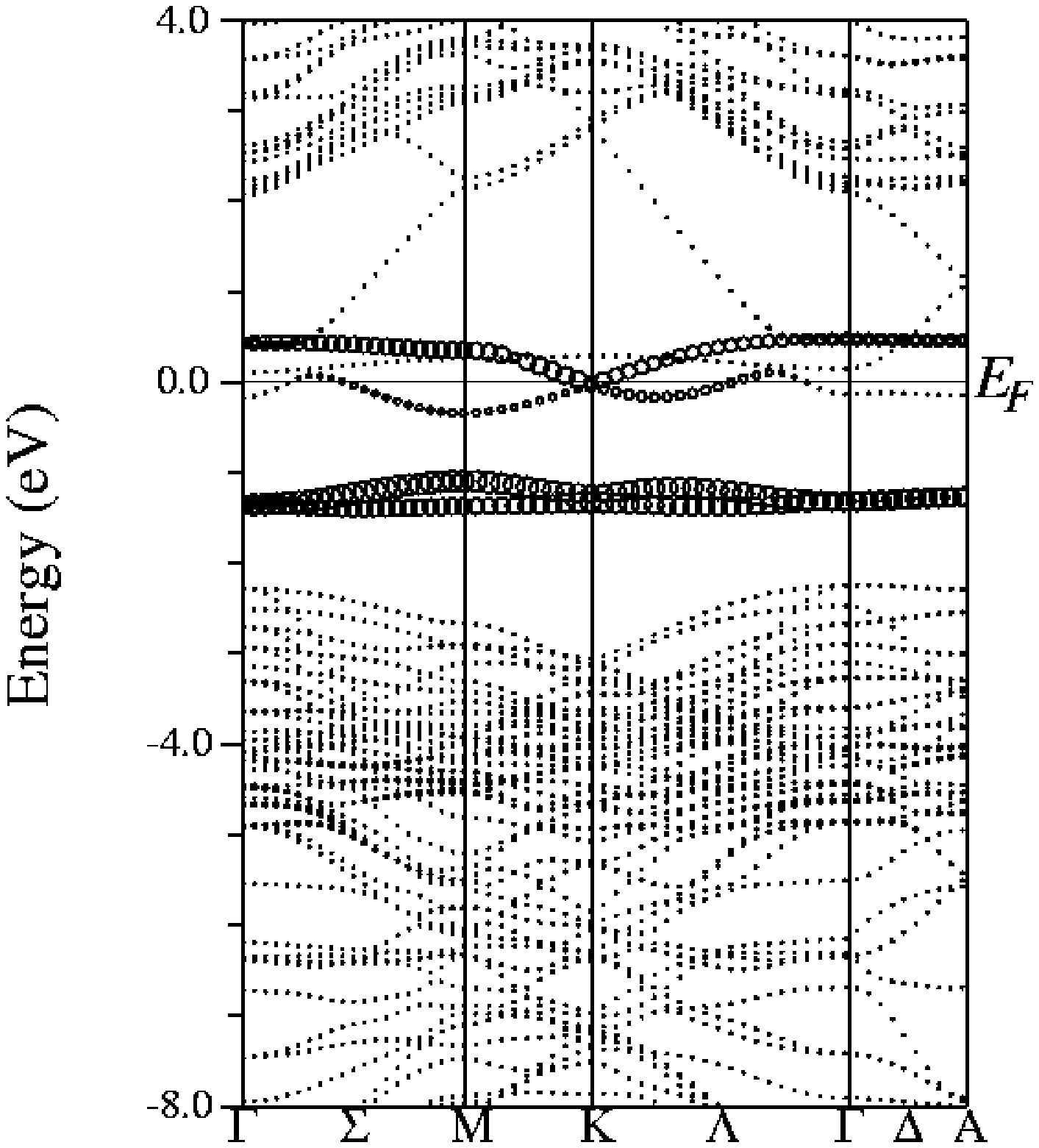}
\hskip 0.2in
\epsfxsize 2.0in
\epsffile {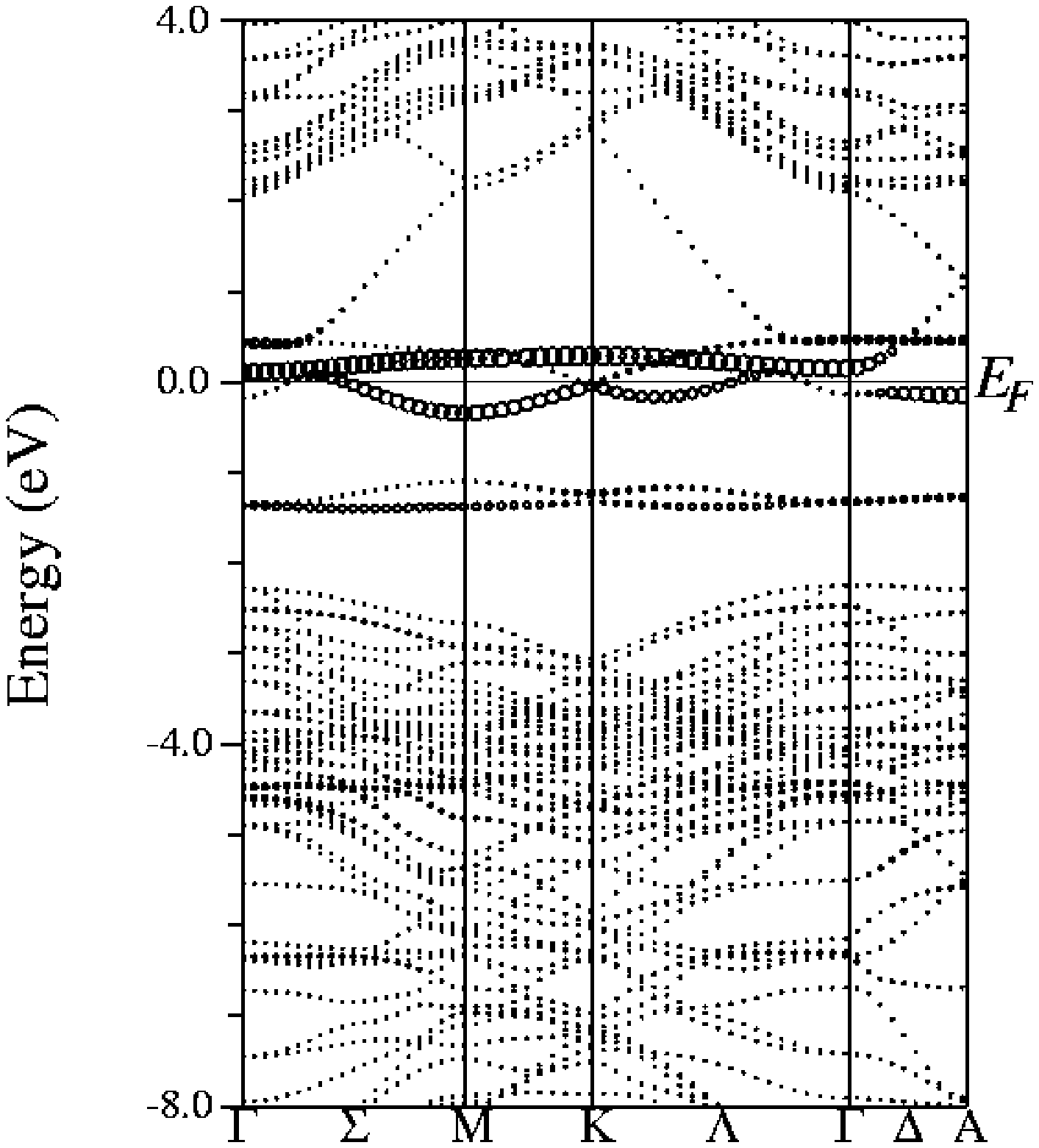}
\vskip 0.02cm
\hskip +2.1in {(a)} \hspace{1.8in} {(b)}
\caption{\label{cret2}Cr-projected majority spin (a) $d_{t_2}$ and (b) $d_{e}$ bands in $Ga_{15}CrN_{16}$.}
\hskip 0.2in
\epsfxsize 1.5in
\end{figure}
The Cr d$_{t_2}$ levels split (figure~\ref{cret2}(a)), out of which two energy levels 
lie below the Fermi level ($\epsilon_{F}$) and are occupied. The third level which is $\sim0.5$eV
above is unoccupied. One of the d$_e$ level is occupied and the other lies just 
above the $\epsilon_{F}$ as seen in figure~\ref{cret2}(b). However, the hybridization of
$d_{t_2}$ and $d_e$ majority spin states is negligible and these levels are well
separated as opposed to Mn doped case. 
As in the case of Mn doping, the Cr minority spin $d$-states are 
above the $\epsilon_{F}$ and so the impurity states at the $\epsilon_{F}$ are $100 \%$ spin
polarized. 

\begin{figure}[h]
\centering
\includegraphics[height=8.5cm,angle=270] {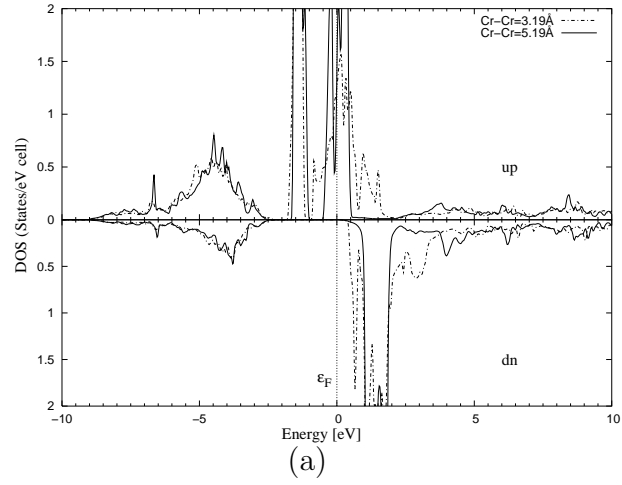}\\
{(a)}\\
\includegraphics[height=8.5cm,angle=270] {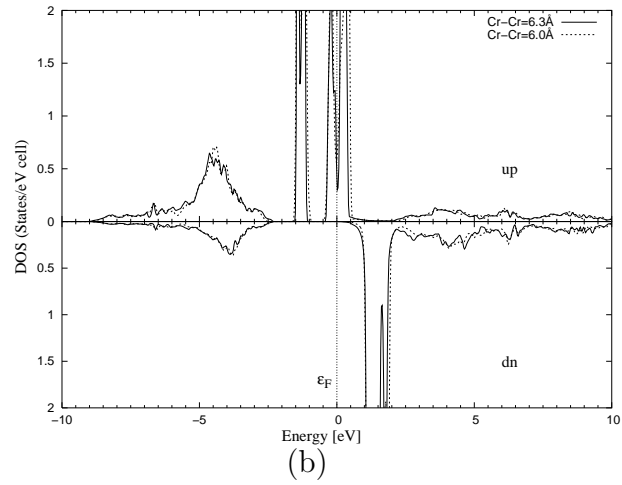}\\
{(b)}
\caption{\label{cr2}Cr-projected $d$-DOS in $Ga_{14}Cr_2N_{16}$ for Cr-Cr separation 
of (a) $3.19$\AA~ and $5.19$\AA~ and (b) $6.0$\AA~ and $6.3$\AA.}
\end{figure}

In the $Ga_{14}Cr_2N_{16}$ system, figure~\ref{cr2} shows that even though
the Cr atoms are substituted at nn sites
there is a gap seen between the split Cr-$d$ levels, unlike in the nn Mn-doped system.
The minority spin conduction band overlaps with the majority spin band for nn Cr case,
as in nn Mn case. For all the geometries of the dopant atoms the system
is half metallic.
\begin{figure}[h]
\centering
\includegraphics[height=5.0cm,width=5cm] {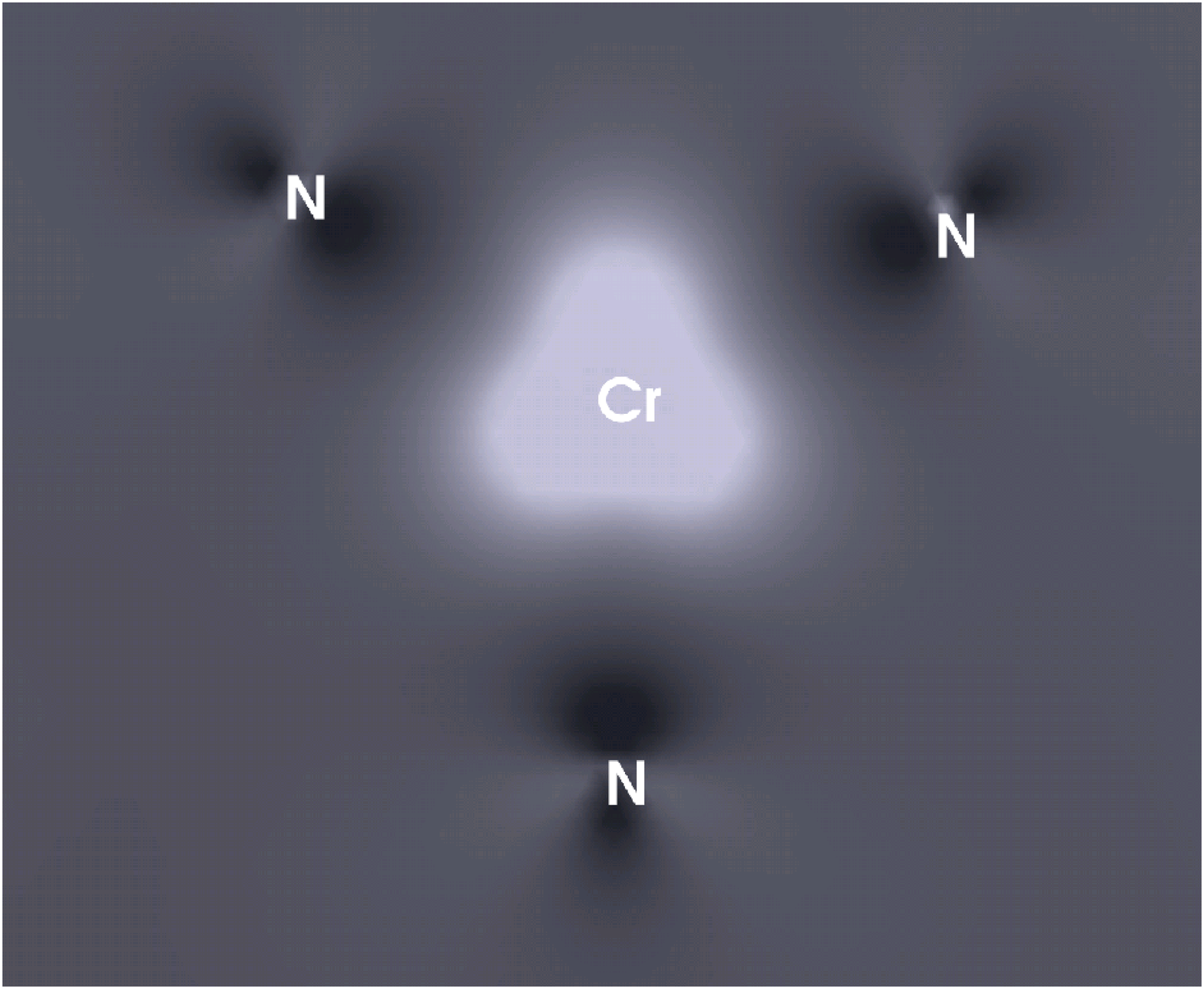}
\includegraphics[height=5.0cm,width=2cm] {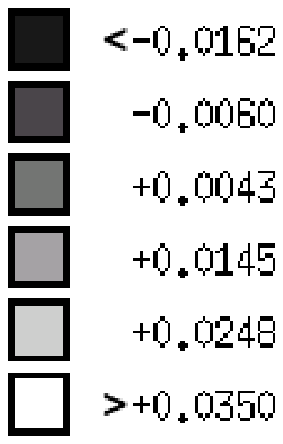}
\caption{\label{crscd} SCD plot in a plane of three nn N atoms around the TM atom in $Ga_{14}CrN_{16}$.}
\end{figure}
The band gap for the minority spin, in case of two Cr substitution at $3.19$\AA~is
$\sim3$eV which is larger than the corresponding Mn case which has a gap of $\sim2.0$eV.
When the TM-TM distance is $5.19$\AA, in the Cr case the gap is $\sim3.0$eV whereas for
Mn substitution it is $\sim2.5$eV. The lowest energy configuration for Cr substitution
occurs for Cr-Cr distance of $3.19$\AA~ and for all the geometries studied the
FM configuration of TM atoms has lower energy compared to AFM configuration. The magnetic moments at
the Cr site in various geometries is as shown in table~\ref{tab2} and it is seen that 
the magnetic moment does not show much variation depending on the distance, indicating
that the direct interaction between the Cr atoms is minimal.

\begin{table}[h]
\caption{\label{tab2}Magnetic moment($\mu_B$) in Cr-doped systems.
Total$_{mom}$ indicates mag. mom./unit cell, dopant$_{mom}$ is the
mag. mom. at dopant site and N$_{mom}$ is the average mag. mom. at
nn N sites.}
%\vskip 0.5in
%\begin{ruledtabular}
\begin{indented}
\item[]\begin{tabular}{@{}llll}
%\begin{tabular}{c c c c}
\br
System & Total$_{mom}$ & Dopant$_{mom}$ & N$_{mom}$\\
%       & mom.($\mu_B$) & at dopant site($\mu_B$) & at nn-N site($\mu_B$)\\
\hline
$Ga_{15}CrN_{16}$     & 3.00 & 2.47 & -0.025\\
\hline
$Ga_{15}Cr_{2}N_{16}$ & 6.00 & 2.48 & -0.031\\
Cr-Cr  = $3.19$ \AA &   &  & \\
\hline
$Ga_{15}Cr_{2}N_{16}$ & 6.00 & 2.47 & -0.022\\
Cr-Cr  = $4.5$ \AA &   &  & \\
\hline
$Ga_{15}Cr_{2}N_{16}$ & 6.00 & 2.48 & -0.025\\
Cr-Cr  = $5.19$ \AA &   &  & \\
\hline
$Ga_{15}Cr_{2}N_{16}$ & 6.00 & 2.48 & -0.026\\
Cr-Cr  = $6.0$ \AA &   &  & \\
\br
\end{tabular}
\end{indented}
\end{table}
Magnetic moments observed at the various dopant sites from our 
calculation are shown in the table~\ref{tab2}.
The total magnetic moment per unit cell per Mn atom is $4\mu_B$ and the average magnetic
moment on the nn N atoms in case of Mn-doping is parallel to the Mn-moment. This can be
understood as penetration of the spin-polarized Mn states to the neighboring host which does
not have any of its own states in the gap region.
The magnetic moment per unit cell per Cr atom is $3\mu_B$. The average magnetic moment
on the nn N atom is anti-parallel to Cr-moment. The difference in the 
orientation of the average magnetic
moment on nn N atoms of Mn and Cr is due to the difference in the $p$ DOS of the nn N along
the z axis (figure not shown here) compared to the nn N atoms lying 
in the xy plane above the TM atoms.
There is not much variation of magnetic moment with increase in Cr-Cr distance
and shows a similar trend as Mn doped systems.
The magnitude of average nn-N magnetic moment is greater in case 
of Cr substitution, which contribute one less electron to the hybridized valcen
band. SCD in figure~\ref{crscd}
on all of the nn-N atoms of the single Cr (only 3 nn N atoms shown in figure~\ref{crscd}) 
doped system is negative thus showing that the TM atom
and the nn-N are anti-ferromagnetically coupled.
\end{subsection}

\begin{subsection}{Estimation of $T_c$}
We have predicted the $T_c$ for the DMS based on Ga$_{16}$Mn$_2$N$_{16}$ and 
Ga$_{16}$Cr$_2$N$_{16}$ considering the mean field approximation.
\begin{figure}[h]
\centering
\includegraphics[height=8.5cm,angle=270] {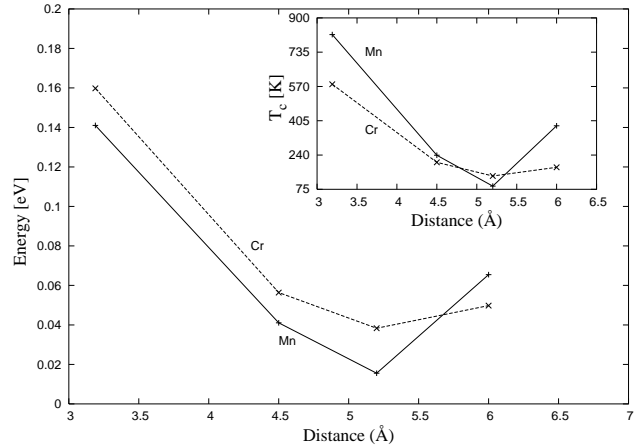}
\caption{\label{evsd} $\Delta E = E_{AFM} - E_{FM}$ for Mn and Cr doping. Inset shows the
mean field $T_c$ variation with distance between dopants.} 
\end{figure}
Figure~\ref{evsd} shows the $\Delta E = E_{AFM} - E_{FM}$ for the Mn/Cr doped systems,
where $E_{AFM}$ is the total energy for the antiferromagnetic
(AFM) configuration and $E_{FM}$ is the total energy for the ferromagnetic (FM) configuration.
Observed variation  $\Delta E$ vs distance for Mn substituted DMS agrees with the
one reported by Sanyal~\cite{mirb}.
$\Delta E$ is a measure of the exchange interaction in the system. 
Highest $\Delta E$ is seen for the case where the TM atoms are substituted as near neighbors,
signifying larger overlap of the magnetic impurity orbitals.
For Mn doping at nn $\Delta E \sim0.14$eV, this compares well with the value calculated for
dimer substitution by Uspenskii {\it et. al }~\cite{uspen} 
which compared qualitatively with the high $T_c = 940$K measured ~\cite{sono}. 
As for the identical Cr case $\Delta E \sim0.16eV$ also compares well with 
the observed $T_c$ but is lower than the $900$K observed by Liu 
and co-workers ~\cite{liu}. From figure~\ref{evsd}
it also emerges that the exchange interaction decreases sharply as the distance
between the TM atoms increases. Thus the exchange interaction is short range and could be
interpreted as the double exchange mechanism.
\end{subsection}
\end{section}

\begin{section}{Summary and conclusions}
We have analyzed the electronic structure of GaN doped with TM Mn and Cr 
with $6.25 \%$ and $12.5 \%$ doping for various
possible geometries to replicate the situation where the TM atoms would appear either 
to cluster or be separated. The self consistent FP-LAPW
calculations predict half metallic state for 
$6.25 \%$ as well as $12.5 \%$ doping. 
Comparing the total energies of the FM and AFM configurations
for $12.5 \%$ doping, the FM state is found to be lower in energy 
and is predicted to be
the preferred state. On-site magnetic moment at the TM site shows insignificant variation
with distance between the dopants. The near neighbor
N atoms contribute to the states in energy gap of the 
semiconductor due to the influence of the TM atoms. Average magnetic moment at nn N site
is parallel to the Mn magnetic moment where as it is anti-parallel to the Cr atoms. We
observe that both the systems with nn substitution of Mn/Cr atom would show high
$T_c$. The energy gap between the minority spin band in Mn is $\sim 1eV$ lower than in
Cr doped system and we think this could be an important factor in determining a suitable
system. But since the magnetic moment at Mn site is higher than Cr, it would be of interest
to study mixed systems of Mn and Cr to incorporate the salient features of both TM atoms.
\end{section}
\ack{We thank BRNS, DAE, Govt. of India for financial support under BARC-UoP 
Collaborative Programme. AK acknowlegdes financial support from DST, Govt. of
India and UGC, Govt. of India. We acknowledge CMS, University of Pune for use
of HPC facility.}


\section*{References}
\begin{thebibliography}{20}

%\bibitem{r1} H. Ohno, Science {\bf 281}, 951-956 (1998), and references therein.

\bibitem{sato} Sato H., Katayama-Yoshida H., 2001 {\it Jpn. J. Appl. Phys.} 
{\bf 40} L485 

\bibitem{cr1} Hashimoto M., Zhou Y., Kanamura M., Asahi H. 2002 {\it Solid State Comm.}
{\bf 122} 37

\bibitem{cr2} Wang J., Chen P., Guo X., Li Z., Lu W. 2005 {\it J. Crys. Growth}
{\bf 275} 393

\bibitem{take} Takeuchi T., Harada Y., Tokushima T., Taguchi M., Takata Y.,
Chainani A., Kim J. J., Makino H., Yao T., Yamamoto T., Tsukamoto T.,
Shin S., Kobayashi K. 2004 {\it Phys. Rev. B} {\bf 70} 245323

\bibitem{liu} Liu H. X., Wu S. Y., Singh R. K., Gu L., Dilley N. R., Montes L.,
Simmonds M. B. 2004 {\it Appl. Phys. Lett.} {\bf 85} 4076

\bibitem{diet} Dietl T., Ohno H., Matsukura F., Cibert J., Ferrand D. 2000
{\it Science} {\bf 287} 1019

\bibitem{reed} Reed M. L., El-Masry N. A., Stadelmaier H. H., Ritums M. K., Reed M. J.,
Parker C. A., Roberts J. C., Bedair S. M. 2001 {\it Appl. Phys. Lett.} {\bf 79} 3473

\bibitem{thal} Thaler G. T., Overberg M. E., Gila B., Frazier R., Abernathy C. R., Pearton S. J.,
Lee J. S., Lee S. Y., Park Y. D., Khim Z. G., Kim J. and Ren F.
2002 {\it Appl. Phys. Lett.} {\bf 80} 3964

\bibitem{sono} Sonoda S., Shimizu S., Sasaki T., Yamamoto Y. and Hori H. 
2002 {\it J. Cryst. Growth} {\bf 237-239} 1358

\bibitem{ando} Ando K., 2003 {\it Appl. Phys. Lett.} {\bf 82} 100 

\bibitem{zaja} Zajac M., Gosk J., Kaminska M., Twardowski A., Szyszko T. and Podsiadlo S. 
2001 {\it Appl. Phys. Lett.} {\bf 79} 2432

\bibitem{uspen} Uspenskii Y., E. Kulatov, H. Mariette, H. Nakayama, H. Ohta
J. Magn. Magn. Mater. {\bf 258-259}, 248-250 (2003).

\bibitem{mirb} Sanyal B. and Mirbt S. 2005 {\it J. M. M. M.} {\bf 290-291} 1408

\bibitem{reib} Raibiger H., Ayuela A. and Nieminen R. M. 2004 {\it J. Phys. Condens. Matter} 
{\bf 16} L457-L462

\bibitem{wien} P. Blaha, K. Schwarz, G. K. H. Madsen, D. Kvasnicka and 
J. Luitz, {\it An augmented plane wave plus local orbitals program for
calculating crystal properties}, Vienna Univ. of Technology,
Austria (2001) ISBN 3-950131-1-2

\bibitem{pbe96} Perdew J. P., Burke K. and Ernzerhof M. 1996 {\it Phys. Rev. Lett.}
{\bf 77} 3865

\bibitem{kron} Kronik L., Jain M. and Chelikowsky R. 2002 {\it Phys. Rev. B}
{\bf 66} 041203

\bibitem{gp1} Das G. P., Rao B. K. and Jena P. 2003
{\it Phys. Rev. B} {\bf 68}, 35207

\bibitem{gp2} Das G. P., Rao B. K. and Jena P. 2004
{\it Phys. Rev. B} {\bf 69}, 214422

%\bibitem{mirb1} Sanyal B., Bengone O. and Mirbt S. 2003 
%{\it Phys. Rev. B} {\bf 68} 205210
%
%
%\bibitem{r8} S. E. Park, H. H. Lee, Y. C. Cho, S. Y. Jeong, C. R. Cho,
%and S. Cho, Appl. Phys. Lett. {\bf 80}, 4187 (2002).
%
%\bibitem{r10} J. S. Lee, J. D. Lim, Z. G. Khim, Y. D. Park, S. J. Pearton, 
%S. N. G. Chu, J. Appl. Phys. {\bf 93}, 4512 (2003).
%
%\bibitem{r16}M. Hashimoto, Y. -K. Zhou, M. Kanamura, and H. Asahi, 
%Solid State Commun. {\bf 122}, 37 (2002).
%
%\bibitem{r12} G. A. Prinz, Science {\bf 282}, 1660-1663 (1998).
%
%\bibitem{r17} B. K. Rao, P. Jena,
%Phys. Rev. Lett. {\bf 89}, 185504-1 (2002).
%
%\bibitem{r13} H. Ohno, D. Chiba, F. Matsukura, T. Omiya, T. Dietl et al.,
%Nature, Vol {\bf 408}, 944 (2000).
%
%\bibitem{r5} T. Sasaki, S. Sonada, Y. Yamamoto, K. Suga, S. Shimizu, K. Kindo 
%and H. Hori, J. Appl. Phys. {\bf 91}, 7911 (2002).
%
%\bibitem{r14} W. Xie and B. Liu, J. Appl. Phys. {\bf 96}, 3569 (2004).
%
%\bibitem{r15} P. Blaha, K. Schwarz, P. Sorantin, and S. B. Trickey,
%Comput. Phys. Commun., {\bf 59}, (1990) 399-415.
%
\end{thebibliography}
\end{document}